\documentclass[preprint,pra,superscriptaddress,showpacs,amsfonts,amsmath,floatfix]{revtex4}

\usepackage{graphicx}
\usepackage{epstopdf}
\usepackage{amssymb}
\usepackage{amsmath}
\usepackage{xspace}
\usepackage{color}

\begin{document}

\title{Quantum theory of a two-mode open-cavity laser}

\author{V. Eremeev}
\affiliation{Facultad de F\'{i}sica, Pontificia Universidad Cat\'{o}lica de Chile, Casilla 306, Santiago 22, Chile}

\author{S.E. Skipetrov}
\affiliation{Univ. Grenoble 1/CNRS, LPMMC UMR 5493, 25 rue des Martyrs, Maison des Magist\`{e}res, 38042 Grenoble, France}

\author{M. Orszag}
\affiliation{Facultad de F\'{i}sica, Pontificia Universidad Cat\'{o}lica de Chile, Av. Vicu\~{n}a Mackenna 4860, Santiago, Chile}

\begin{abstract}
We develop the quantum theory of an open-cavity laser assuming that only two modes compete for gain. We show that the modes interact to build up a composite mode that becomes the lasing mode when pumping exceeds a threshold. This composite mode exhibits all the features of a typical laser mode, whereas its precise behavior depends explicitly on the openness of the cavity. We approach the problem by using the density-matrix formalism and derive the master equation for the light field.
Our results are of particular interest in the context of random laser systems.
\end{abstract}

\maketitle

\section{Introduction}

Small size, complex structure and extreme openness or complete absence of the cavity are characteristic features of a number of ``exotic'' laser systems that have attracted the attention of physicists in recent years \cite{nano10}. Examples of lasers that fall into this category are chaotic microcavity \cite{gmachl98,shinohara10} and random \cite{cao03,cao05,wiersma08} lasers. These systems are quite different from ``traditional'' cavity lasers composed of an amplifying medium in a high quality-factor cavity \cite{siegman86,Scully}. From the theoretical point of view, the very strong coupling to the external world requires a special treatment that is different from what can be found in standard laser textbooks \cite{siegman86,Scully}. A semiclassical model of lasing in open complex or random media was developed by T\"{u}reci \emph{et al.} \cite{tureci06}. This theory was successfully applied to understand lasing in random media \cite{tureci08}. To develop the quantum theory of an open-cavity laser, one starts by facing the problem of  quantization of the electromagnetic field in a space that cannot be separated into ``system'' and ``bath'' parts unambiguously. This problem was solved by Hackenbroich \emph{et al.} \cite{hackenbroich02} who also put forward Langevin and master equations to describe the dynamics of modes in open resonators \cite{hackenbroich03}. Hackenbroich also derived Heisenberg-Langevin equations for an open-cavity laser, which, however, he analyzed only in the semiclassical approximation \cite{hackenbroich05}.
A related problem of light emission by an atom in a lossy cavity was also considered by Di Fidio \textit{et al.} \cite{di fidio} who, however, considered a single-mode field and hence didn't discuss additional features that arise from the coupling between modes due to the openness of the system.

In the present paper we use a combination of the quantization procedure of Refs.\ \cite{hackenbroich02} with the standard density operator approach \cite{Scully} to develop the full quantum theory of an open-cavity laser. We compute and analyze the lasing threshold, the photon statistics (both below and above the threshold), as well as the emission linewidth of a laser that has no well-defined cavity, assuming that only two modes compete for gain. This simple two-mode model allows us to capture some of the essential features of cooperative mode dynamics that seems to determine the behavior of the system. We compare our results with those known from the standard laser theory \cite{Scully} and highlight common features as well as important differences. It is worthwhile to note that a different master equation for a random laser was previously proposed by Florescu and John \cite{florescu04}. These authors considered the random laser as a collection of low quality-factor cavities, coupled by random photon diffusion. In contrast to this work, our approach has the advantage of not relying on any particular model of wave transport, as well as being based on a well-defined quantization procedure for the electromagnetic field and a fully quantum model for the atoms providing amplification.

\section{Master equation for the reduced density operator of the electromagnetic field}

Let us start by considering an ensemble of two-level atoms interacting with the electromagnetic field. We divide the modes of the electromagnetic field into those that belong to the system ``atoms + field'' (A + F) and those that constitute the ``bath''. In the density operator approach, the system A + F is described by the density operator $\hat{\rho}(t)$. The reduced density operator $\hat{\rho}_F(t)$ describing the electromagnetic field (F) is obtained by tracing over the atomic (A) degrees of freedom: $\hat{\rho}_F(t) = \mathrm{Tr}_A \hat{\rho}(t)$. The dynamics of the laser is due to the competition between gain (due to the interaction of the field with atoms) and loss (due to the coupling of the system A + F to the bath), which can be expressed in the form of the following master equation:
\begin{equation}
\dot{\hat{\rho}}_{F}=\hat{L}^{\text{(gain)}}\hat{\rho}_{F}+
\hat{L}^{\text{(loss)}}\hat{\rho}_{F}.
\label{eq:ME}
\end{equation}
Here the super-operators $\hat{L}^{\text{(gain)}}$ and $\hat{L}^{\text{(loss)}}$ describe the gain and the loss, respectively.

Equation (\ref{eq:ME}) is quite formal and can be written for any quantum system interacting with environment. Let us now give expressions for the two terms on the r.h.s. of Eq.\ (\ref{eq:ME}) in the case of an open-cavity laser. A way to deal with the second term $\hat{L}^{\text{(loss)}}\hat{\rho}_{F}$ was proposed in a series of papers by Hackenbroich \emph{et al.} \cite{hackenbroich02,hackenbroich03}. The idea is to separate the physical space $\mathbb{R}^3$  into two subspaces and to quantize the field in terms of the modes $a$ and $b$ of these subspaces. In the context of the laser system considered here, it is natural to choose the first subspace such that it contains all the atoms and has a finite volume. The discrete modes of the first subspace will constitute our sub-system F, whereas the modes of the second subspace will make up the bath. An equation for the density matrix of the sub-system F is derived by tracing over the degrees of freedom corresponding to the modes that belong to the bath. This yields \cite{hackenbroich03}
\begin{equation}
\hat{L}^{\text{(loss)}}\hat{\rho}_{F}=
\sum_{\lambda,\lambda'}
\gamma_{\lambda\lambda'}
\left(2\hat{a}_{\lambda'}\hat{\rho}_{F}\hat{a}_{\lambda}^{\dagger}
-\hat{\rho}_{F}\hat{a}_{\lambda}^{\dagger}\hat{a}_{\lambda'}-
\hat{a}_{\lambda}^{\dagger}\hat{a}_{\lambda'}\hat{\rho}_{F}\right).
\label{eq:losses rho}
\end{equation}
Here $\hat{a}_{\lambda}$ and $\hat{a}_{\lambda}^{\dagger}$ are annihilation and creation operators corresponding to the modes of the sub-system F, and the coefficients $\gamma_{\lambda\lambda'}$ depend on the precise geometry of the system. These coefficients were calculated for a number of particular open cavities in Ref.\ \cite{viviescas04} but may be difficult to obtain in the general case. In a random laser system, they may be treated as random variables \cite{hackenbroich05}.

The essential difference between Eq.\ (\ref{eq:losses rho}) and the analogous equation of the standard laser theory \cite{Scully} is that the damping matrix $\gamma$ is not diagonal. This shows that the openness of the system not only leads to losses described by the diagonal elements of $\gamma$, but also induces coupling between different modes. The strength of the coupling is given by the off-diagonal elements of the damping matrix $\gamma$.

Let us now turn to the first term in Eq.\ (\ref{eq:ME}). It is not specific for the open-cavity laser, so that we will follow standard approaches to derive an explicit expression for it \cite{Scully,bergou89,orszag08}. As the first step, we consider the Jaynes-Cummings Hamiltonian for an atom interacting with the electromagnetic field
(we set $\hbar=1$ in the following) \cite{jaynes63}:
\begin{equation}
\hat{H}=\frac{\omega_{a}}{2}\hat{\sigma}_{z}+
\sum_{\lambda}\omega_{\lambda}\hat{a}_{\lambda}^{\dagger}
\hat{a}_{\lambda}+\sum_{\lambda}(g_{\lambda}\hat{\sigma}^{\dagger}
\hat{a}_{\lambda}+\mathrm{h.c.}).
\label{eq: J-C ham}
\end{equation}
Here $\omega_a$ is the frequency of the atomic transition, $\hat{\sigma}^{\dagger} = |e\rangle \langle g|$ and $\hat{\sigma}_z = |e\rangle \langle e| - |g\rangle \langle g|$ are the atomic raising and inversion operators, respectively, the states $|g\rangle$ and $|e\rangle$ are the ground  and excited states of the two-level atom, $\omega_{\lambda}$ are the frequencies of the modes of the field, and the coefficients $g_{\lambda}$ describe the coupling between the atom and the mode $\lambda$ of the field.
It is convenient to introduce a reference frequency $\bar{\omega}$ and the detuning parameters $\delta = \omega_{a}-\bar{\omega}$ and
$\Delta_{\lambda} = \omega_{\lambda}-\bar{\omega}$ to write
\begin{eqnarray}
\hat{H} &=& \frac{\bar{\omega}}{2}\hat{\sigma}_{z}+
\bar{\omega}\sum_{\lambda}\hat{a}_{\lambda}^{\dagger}
\hat{a}_{\lambda}+\frac{\delta}{2}\hat{\sigma}_{z}+
\sum_{\lambda}\Delta_{\lambda}\hat{a}_{\lambda}^{\dagger}
\hat{a}_{\lambda}+\sum_{\lambda}(g_{\lambda}\hat{\sigma}^{\dagger}
\hat{a}_{\lambda}+\textrm{h.c.})\nonumber \\
&=& \hat{H}_{0} + \hat{V},
\end{eqnarray}
where $\hat{H}_{0} = \bar{\omega}\hat{\sigma}_{z}/2+\bar{\omega}\sum_{\lambda}
\hat{a}_{\lambda}^{\dagger}\hat{a}_{\lambda}$ and
$\hat{V} = \delta\hat{\sigma}_{z}/2+\sum_{\lambda}\Delta_{\lambda}
\hat{a}_{\lambda}^{\dagger}\hat{a}_{\lambda}+\sum_{\lambda}
(g_{\lambda}\hat{\sigma}^{\dagger}\hat{a}_{\lambda}+\textrm{h.c.})$.
Because $\hat{H}_0$ and $\hat{V}$ commute, $[\hat{H}_{0}, \hat{V}] = 0$, we will work in the interaction picture where the dynamics of the system is governed by $\hat{V}$. In this picture, the time evolution of the density operator is given by the evolution operator $\hat{U}(t)=\exp[-i\hat{V}t]$:
\begin{equation}
\hat{\rho}(t)=\hat{U}(t)\hat{\rho}(0)\hat{U}^{\dagger}(t),
\label{eq:ro evol}
\end{equation}
where
\begin{equation}
\hat{V}=\begin{pmatrix}\delta/2+\sum_{\lambda}\Delta_{\lambda} \hat{a}_{\lambda}^{\dagger}\hat{a}_{\lambda} & \sum_{\lambda}g_{\lambda}\hat{a}_{\lambda}\\
\sum_{\lambda}g_{\lambda}^{*}\hat{a}_{\lambda}^{\dagger} & -\delta/2+\sum_{\lambda}\Delta_{\lambda}\hat{a}_{\lambda}^{\dagger}\hat{a}_{\lambda}\end{pmatrix}.\end{equation}
We will now restrict our consideration to the situations when the frequencies $\omega_{\lambda}$ of the modes are close to $\bar{\omega}$, such that $\Delta_{\lambda} \ll g_{\lambda}$, $\delta$, and will proceed by setting $\Delta_{\lambda} = 0$:
\begin{equation}
\hat{V} = \begin{pmatrix}\delta/2 & g\hat{A}\\
g\hat{A}^{\dagger} & -\delta/2\end{pmatrix}.
\end{equation}
Here $g\hat{A}=\sum_{\lambda}g_{\lambda}\hat{a}_{\lambda}$ and $g=(\sum_{\lambda}g_{\lambda}^{2})^{1/2}$. The newly defined operator ${\hat A}$ obeys the standard bosonic commutation relation $[\hat{A}, \hat{A}^{\dagger} ] = 1$. 
After some algebra, with the help of the operators $\hat{\varphi} = g [\hat{A}\hat{A}^{\dagger}+(\delta/2g)^{2}]^{1/2}$ and $\hat{\phi} = g [\hat{A}^{\dagger}\hat{A}+(\delta/2g)^{2}]^{1/2}$,  the evolution operator reads
\begin{equation}
\hat{U}(t)=\begin{pmatrix} \cos[\hat{\varphi}t] - i\left(\delta/2\right)\sin[\hat{\varphi}t]/\hat{\varphi} & -i \sin[\hat{\varphi}t]/\hat{\varphi} \hat{A}\\
-i\hat{A}^{\dagger}\sin[\hat{\varphi}t]/\hat{\varphi} & \cos[\hat{\phi}t] + i\left(\delta/2\right)\sin[\hat{\phi}t]/\hat{\phi} \end{pmatrix}
\label{eq:evol oper}
\end{equation}

We now assume that at the initial time $t = 0$,
$\hat{\rho}(0)=\hat{\rho}_{F}(0)\otimes\hat{\rho}_{A}(0)$, with $\hat{\rho}_{A}$ being the density operator of the atom, and that the atom is in its upper state: $\hat{\rho}_{A}(0) = |2\rangle\langle2| = (\hat{\sigma}_{z}+1)/2$. The density matrix of the full system ``atom + field'' at $t = 0$ is then
\begin{equation}
\hat{\rho}(0)=\begin{pmatrix}\hat{\rho}_{F}(0) & 0\\
0 & 0\end{pmatrix}.
\end{equation}
Considering $\hat{\Phi}_{\pm} = \cos[\hat{\varphi}t] \pm i\left(\delta/2\right)\sin[\hat{\varphi}t]/\hat{\varphi}$,
equations (\ref{eq:ro evol}) and (\ref{eq:evol oper}) allow us to compute the density operator at arbitrary time as
\begin{equation}
\hat{\rho}(t)=\begin{pmatrix}\hat{\Phi}_{-}
\hat{\rho}_{F}(0)\hat{\Phi}_{+} & ig\hat{\Phi}_{-}\hat{\rho}_{F}(0)
\sin\left[\hat{\varphi}t\right]/\hat{\varphi}\hat{A}\\
\, & \,\\
-ig\hat{A}^{\dagger}\sin\left[\hat{\varphi}t\right]/
\hat{\varphi}\hat{\rho}_{F}(0)\hat{\Phi}_{+}\;\; & g^{2}\hat{A}^{\dagger}\sin\left[\hat{\varphi}t\right]/\hat{\varphi}\hat{\rho}_{F}(0)\sin\left[\hat{\varphi}t\right]/\hat{\varphi}\hat{A}\end{pmatrix}.\end{equation}
Finally, the reduced density operator $\hat{\rho}_{F}(t)=\text{Tr}_{A}\hat{\rho}(t)$ is
\begin{eqnarray}
\hat{\rho}_{F}(t) &=& \hat{\Phi}_{-}\hat{\rho}_{F}(0)\hat{\Phi}_{+}+
g^{2}\hat{A}^{\dagger}\sin\left[\hat{\varphi}t
\right]/\hat{\varphi}\hat{\rho}_{F}(0)\sin\left[
\hat{\varphi}t\right]/\hat{\varphi}\hat{A}
\nonumber \\
&=& \hat{\Lambda}(t) \hat{\rho}_F(0),
\label{rhof}
\end{eqnarray}
where we defined a super-operator $\hat{\Lambda}(t)$.

Equation (\ref{rhof}) yields the evolution of the reduced density operator of the electromagnetic field interacting with a two-level atom which is initially in the excited state. This is clearly insufficient to describe laser emission. The first lacking ingredient stems from the fact that we want to describe an ensemble of many atoms, not just a single atom. To make use of Eq.\ (\ref{rhof}) in the case of many atoms, we assume that (i) the atoms interact with the field one after another, in sequence, and not all at a time, and (ii) the time of interaction of a given atom with the field $\tau$ is much shorter than the typical time $t$ at which the evolution of the field is calculated. The density matrix of the field after a time $t \gg \tau$ during which the field interacted with $k$ atoms will be then equal to $\hat{\rho}_F^{(k)}(t) = \hat{\Lambda}^k(\tau) \hat{\rho}_F(0)$ \cite{Scully,orszag08}. We will now introduce the second important ingredient of the laser system --- the pump. To model the pump in the framework of our two-level atom model, we assume that atoms are transferred to the excited state at a rate $r$ by some external mechanism (for example, via additional atomic levels that are not included in our model explicitly) and that the probability for $k$ atoms to get excited during a time $\Delta t$ is $P(k) = C_{Kk} p^{k} (1-p)^{K-k}$, where
$C_{Kk} = K!/k!(K-k)!$, $p$ is the probability for a given atom to be in the excited state, and $K$ is the total number of atoms that can potentially participate in the lasing process (i.e., $0 \leq k \leq K$). The average number of excited atoms is $\langle k \rangle = pK = r \Delta t$. The parameter $p$ describes statistics of pumping, with the limit $p\rightarrow0$ (that we will consider from here on) corresponding to random pumping and the limit $p\rightarrow1$ corresponding to a uniform (regular) pumping \cite{bergou89, Yamamoto}.

The density operator averaged over $k$ is \cite{bergou89,orszag08}
\begin{equation}
\hat{\rho}_{F}(t)=\sum_{k=0}^{K} P(k)
\hat{\rho}_{F}^{(k)}(t)=\left\{1 + p[\hat{\Lambda}(\tau)-1] \right\}^{K} \hat{\rho}_{F}(0).
\label{eq:ro probab}
\end{equation}

To obtain a dynamic equation for $\hat{\rho}_F(t)$, we take the derivative of Eq.\ (\ref{eq:ro probab}) with respect to time and expand the result in series in $p (\hat{\Lambda} - 1)$:
\begin{eqnarray}
\dot{\hat{\rho}}_{F}(t) &=& \frac{r}{p}
\ln\left\{1+p[\hat{\Lambda}(\tau)-
1]\right\}\hat{\rho}_{F}(t)
\nonumber \\
&\simeq& r [\hat{\Lambda}(\tau)-1]\hat{\rho}_{F}(t)
-\frac{r p}{2}[\hat{\Lambda}(\tau)-1]^{2}\hat{\rho}_{F}(t).
\label{eq:deriv2 rho}
\end{eqnarray}
Finally, we now take into account the fact that the time of interaction of a given atom with the field $\tau$ is, in fact, a random variable. $\tau$ is finite due to the possible decay of the excited state without coupling to the modes of the electromagnetic field that make part of our sub-system F. This decay may be due, for example, to transitions involving additional atomic levels (with or without emission of a photon), not included in our model. With $\Gamma$ being the rate of such transitions, the statistical distribution of $\tau$ is $P(\tau) = \Gamma \exp(-\Gamma \tau)$. By averaging Eq.\ (\ref{eq:deriv2 rho}) over this distribution, we obtain
\begin{equation}
\hat{L}^{\text{(gain)}}\hat{\rho}_{F} = r \int_{0}^{\infty} d\tau \Gamma
\exp(-\Gamma \tau) \left\{ [\hat{\Lambda}(\tau)-1]-\frac{p}{2}
[\hat{\Lambda}(\tau)-1]^{2}\right\}\hat{\rho}_{F}(t).
\label{eq:rho gain fin}
\end{equation}

Equations (\ref{eq:losses rho}) and (\ref{eq:rho gain fin}) provide explicit expressions for the two terms on the r.h.s. of Eq.\ (\ref{eq:ME}). It is worthwhile to note that Eq.\ (\ref{eq:losses rho}) was derived in the Schr\"{o}dinger picture, whereas Eq.\ (\ref{eq:rho gain fin}) --- in the interaction picture. In the interaction picture, the general form of Eq.\ (\ref{eq:losses rho}) remains unchanged, except for the damping matrix that has to be transformed accordingly. In the present paper we will not use any particular model for this matrix but will rather treat it as a free parameter, having in mind that in a random laser, for example, it is a random matrix.

\section{Two-mode model}

Under general conditions, many modes may coexist and compete for gain in an open-cavity laser. The off-diagonal elements of the damping matrix $\gamma$ couple the modes and make the single-mode regime hardly realizable. It is worthwhile to note that this coupling is different from the coupling via interaction with the atomic subsystem which is always present and hardly depends on the type of the cavity under consideration. To analyze the interaction between modes, we consider a model in which only two modes are taken into account. This simple situation often allows for important insights into the dynamics of laser systems, as, for example, it was the case for the quantum-beat or correlated-emission lasers \cite{Bergou}. Our model differs from the previously considered two-mode models (see, e.g., Ref.\ \cite{Bergou} but also Refs.\ \cite{singh80} and \cite{Swain}) by the mode coupling through both the atomic subsystem (i.e., the coupling due to the fact that the modes interact with the same atomic transition) and the \textit{common} bath, whereas only the first type of coupling was considered in Refs.\ \cite{Bergou,singh80,Swain}. Thus, having a common bath for the two modes is essential in our model. The strength of the additional coupling between the modes is given by the off-diagonal elements $\gamma_{12}$ and $\gamma_{21}$ of the damping matrix $\gamma$. To put accent on this new element of the model, we will focus on the dependence of our results on these off-diagonal elements in what follows.
To proceed, we rewrite the master equation (\ref{eq:ME}) assuming the limit of $p\rightarrow0$:
\begin{eqnarray}
\dot{\hat{\rho}} &=& r\int_{0}^{\infty}d\tau\Gamma\exp(
-\Gamma \tau) \left[\hat{\Phi}_{-}\hat{\rho}(t)\hat{\Phi}_{+} +g^{2}\hat{\alpha}^{\dagger}\sin\left[\hat{\varphi}
\tau\right]/\hat{\varphi}\hat{\rho}(t)\sin\left[
\hat{\varphi}\tau\right]/\hat{\varphi}\hat{\alpha}-
\hat{\rho}(t)\right] \nonumber \\
&+& \hat{L}^{\text{(loss)}}\hat{\rho},
\label{eq:ME 2modes}
\end{eqnarray}
where $\hat{\alpha} = (g_{1}\hat{a}_{1}+g_{2}\hat{a}_{2})/g$
and $\hat{\varphi} = g[\hat{\alpha}\hat{\alpha}^{\dagger}+(\delta/2g)^{2}]^{1/2}$.
To lighten the notation, we drop the subscript ``F'' of the reduced density operator and write $\hat{\rho}_F = \hat{\rho}$. The limit $p \rightarrow 0$ corresponds to the Poissonian distribution of the number of excited atoms, $P(k)$, and hence to the realistic case of random pumping by an external source. The loss term in Eq.\ (\ref{eq:ME 2modes}) follows from Eq.\ (\ref{eq:losses rho}):
\begin{eqnarray}
\hat{L}^{\text{(loss)}}\hat{\rho} &=& \gamma_{11}(2\hat{a}_{1}\hat{\rho} \hat{a}_{1}^{\dagger}-\hat{\rho} \hat{a}_{1}^{\dagger}\hat{a}_{1}-\hat{a}_{1}^{\dagger}\hat{a}_{1}\hat{\rho})
+\gamma_{12}(2\hat{a}_{2}\hat{\rho} \hat{a}_{1}^{\dagger}-\hat{\rho} \hat{a}_{1}^{\dagger}\hat{a}_{2}-\hat{a}_{1}^{\dagger}\hat{a}_{2}\hat{\rho})
\nonumber \\
&+& \gamma_{21}(2\hat{a}_{1}\hat{\rho} \hat{a}_{2}^{\dagger}-\hat{\rho} \hat{a}_{2}^{\dagger}\hat{a}_{1}-\hat{a}_{2}^{\dagger}\hat{a}_{1}\hat{\rho})
+\gamma_{22}(2\hat{a}_{2}\hat{\rho} \hat{a}_{2}^{\dagger}-\hat{\rho} \hat{a}_{2}^{\dagger}\hat{a}_{2}-\hat{a}_{2}^{\dagger}\hat{a}_{2}\hat{\rho}).
\label{eq:losses 2modes}
\end{eqnarray}
To solve the master equation (\ref{eq:ME 2modes}), we will work with composite modes $\hat{\alpha}$ defined above and $\hat{\beta} = (g_{2}a_{1}-g_{1}a_{2})/g$, with the commutation relations: $[\hat{\beta},\hat{\beta}^{\dagger}]
= [\hat{\alpha},\hat{\alpha}^{\dagger}] = 1$ and
$[\hat{\alpha},\hat{\beta}^{\dagger}]
= [\hat{\alpha},\hat{\beta}] = 0$ (here we assume that $g_{1,2}$ are real numbers).
The composite modes $\alpha$ and $\beta$ are not due to the phase locking phenomenon but are simply linear superpositions of the bare modes 1 and 2. Using the modes $\alpha$ and $\beta$ instead of $1$ and $2$ simplifies further analysis because only the mode $\alpha$ will be actually excited in the lasing process, as we will see from the following.

To simplify analytical calculations, we will assume $\gamma_{21}=\gamma_{12}$ from here on. This corresponds, for example, to open cavities considered in Ref.\ \cite{viviescas04}. The density operator can be represented in the basis of Fock states $|n_{\alpha}, n_{\beta} \rangle$ as
\begin{equation}
\hat{\rho} = \sum_{{n_{\alpha},n_{\beta}\atop m_{\alpha},m_{\beta}}}\rho_{n_{\alpha},n_{\beta};\, m_{\alpha},m_{\beta}}|n_{\alpha}, n_{\beta}\rangle\langle m_{\alpha}, m_{\beta}|.
\label{eq:Fock st}
\end{equation}
Equations (\ref{eq:ME 2modes}) and (\ref{eq:losses 2modes}) yield an equation for the density matrix
$\rho_{n_{\alpha},n_{\beta};\, m_{\alpha},m_{\beta}}$:
\begin{eqnarray}
&&\dot{\rho}_{n_{\alpha},n_{\beta};\, m_{\alpha},m_{\beta}} =\frac{A\sqrt{n_{\alpha}m_{\alpha}}}{1+\bar{\delta}^{2}+(B/2A)(n_{\alpha}+m_{\alpha})+(B/4A)^{2}(n_{\alpha}-m_{\alpha})^{2}}\rho_{n_{\alpha}-1,n_{\beta};\, m_{\alpha}-1,m_{\beta}}
\nonumber \\
&-&\biggl[\frac{A (n_{\alpha}+m_{\alpha}+2)/2+iA \bar{\delta}(n_{\alpha}-m_{\alpha})/2+B(n_{\alpha}-m_{\alpha})^{2}/8}{1+\bar{\delta}^{2}+(B/2A)(n_{\alpha}+m_{\alpha}+2)+(B/4A)^{2}(n_{\alpha}-m_{\alpha})^{2}}
\nonumber \\
&+&C_{1}(n_{\alpha}+m_{\alpha})/2+C_{2}(n_{\beta}+m_{\beta})/2\biggr]\rho_{n_{\alpha},n_{\beta};\, m_{\alpha},m_{\beta}}
\nonumber \\
&+&C_{1}\sqrt{(n_{\alpha}+1)(m_{\alpha}+1)}\rho_{n_{\alpha}+1,n_{\beta};\, m_{\alpha}+1,m_{\beta}}+C_{2}\sqrt{(n_{\beta}+1)(m_{\beta}+1)}\rho_{n_{\alpha},n_{\beta}+1;\, m_{\alpha},m_{\beta}+1}
\nonumber \\
&+&2C_{3}\sqrt{(n_{\alpha}+1)(m_{\beta}+1)}\rho_{n_{\alpha}+1,n_{\beta};\, m_{\alpha},m_{\beta}+1}+2C_{3}\sqrt{(n_{\beta}+1)(m_{\alpha}+1)}\rho_{n_{\alpha},n_{\beta}+1;\, m_{\alpha}+1,m_{\beta}}
\nonumber \\
&-&C_{3}\sqrt{(m_{\alpha}+1)m_{\beta}}\rho_{n_{\alpha},n_{\beta};\, m_{\alpha}+1,m_{\beta}-1}-C_{3}\sqrt{m_{\alpha}(m_{\beta}+1)}\rho_{n_{\alpha},n_{\beta};\, m_{\alpha}-1,m_{\beta}+1}
\nonumber \\
&-&C_{3}\sqrt{(n_{\alpha}+1)n_{\beta}}\rho_{n_{\alpha}+1,n_{\beta}-1;\, m_{\alpha},m_{\beta}}-C_{3}\sqrt{n_{\alpha}(n_{\beta}+1)}\rho_{n_{\alpha}-1,n_{\beta}+1;\, m_{\alpha},m_{\beta}},
\label{eq:rho nondiag}
\end{eqnarray}
where we defined
$A = 2r(g/\Gamma)^{2}$,
$B = 4(g/\Gamma)^{2} A$,
$\bar{\delta}=\delta/\Gamma$, $C_{1} = 2g^{-2}(\gamma_{11}g_{1}^{2}+2\gamma_{12}g_{1}g_{2}
+\gamma_{22}g_{2}^{2})$,
$C_{2} = 2g^{-2}(\gamma_{11}g_{2}^{2}-2\gamma_{12}g_{1}g_{2}
+\gamma_{22}g_{1}^{2})$ and
$C_{3} = g^{-2} [(\gamma_{11}-\gamma_{22})g_{1}g_{2}+
2\gamma_{12}(g_{2}^{2}-g_{1}^{2})]$.
In order to facilitate the comparison with the standard laser theory, we defined the coefficients $A$ and $B$ in the same way as in the book \cite{Scully} (p. 333). When $\gamma_{12} = 0$ and $\gamma_{11} = \gamma_{22}$, the coefficients $C_1$ and $C_2$ reduce to the coefficient $C$ of this book (p. 255) and $C_3$ vanishes. The coefficient $C_3$ vanishes also for $\gamma_{11} = \gamma_{22}$ and $g_{1} = g_{2} $. This case is somewhat special because it corresponds to two modes with the same losses and the same coupling with atoms. This leads to $C_3 = 0$, but still $C_1 \ne C_2$ for $\gamma_{12} \ne 0$ and hence the problem does not reduce to the case of a high quality-factor cavity.
High quality-factor cavities are described by Eq.\ (\ref{eq:rho nondiag}) with $\gamma_{12} \to 0$. The coupling between the bare modes 1 and 2 then arises uniquely from their interaction with the same atomic transition (and not with the common bath) and the composite modes $\alpha$ and $\beta$ still provide a useful basis for the description of lasing (see, e.g., Ref.\ \cite{Bergou}).

For the diagonal elements of the density matrix $\rho_{n_{\alpha},n_{\beta};\, n_{\alpha},n_{\beta}} = \rho_{n_{\alpha},n_{\beta}}$ and up to the second order in $B/A = 4(g/\Gamma)^2 \ll 1$ we obtain
\begin{eqnarray}
&&\dot{\rho}_{n_{\alpha},n_{\beta}} =\frac{An_{\alpha}}{1+\bar{\delta}^{2}+(B/A)n_{\alpha}}\rho_{n_{\alpha}-1,n_{\beta}}-\biggl[\frac{A(n_{\alpha}+1)}{1+\bar{\delta}^{2}+(B/A)(n_{\alpha}+1)}+C_{1}n_{\alpha}+C_{2}n_{\beta}
\nonumber \\
&-&\frac{2C_{3}^{2}(n_{\alpha}+1)n_{\beta}}{K_{n_{\alpha}+1,\, n_{\beta}}}-\frac{2C_{3}^{2}n_{\alpha}(n_{\beta}+1)}{K_{n_{\alpha},\, n_{\beta}+1}}\biggr]\rho_{n_{\alpha},n_{\beta}}+\frac{8C_{3}^{2}(n_{\alpha}+1)(n_{\beta}+1)}{K_{n_{\alpha}+1,\, n_{\beta}+1}}\rho_{n_{\alpha}+1,n_{\beta}+1}
\nonumber \\
&+& \frac{2C_{3}^{2}(n_{\alpha}+1)n_{\beta}}{K_{n_{\alpha}+1,\, n_{\beta}}}\rho_{n_{\alpha}+1,n_{\beta}-1}+\frac{2C_{3}^{2}n_{\alpha}(n_{\beta}+1)}{K_{n_{\alpha},\, n_{\beta}+1}}\rho_{n_{\alpha}-1,n_{\beta}+1}
\nonumber \\
&+&\left[C_{1}(n_{\alpha}+1)-\frac{4C_{3}^{2}(n_{\alpha}+1)(n_{\beta}+1)}{K_{n_{\alpha}+1,\, n_{\beta}+1}}-\frac{4C_{3}^{2}(n_{\alpha}+1)n_{\beta}}{K_{n_{\alpha}+1,\, n_{\beta}}}\right]\rho_{n_{\alpha}+1,n_{\beta}}
\nonumber \\
&+&\left[C_{2}(n_{\beta}+1)-\frac{4C_{3}^{2}(n_{\alpha}+1)(n_{\beta}+1)}{K_{n_{\alpha}+1,\, n_{\beta}+1}}-\frac{4C_{3}^{2}n_{\alpha}(n_{\beta}+1)}{K_{n_{\alpha},\, n_{\beta}+1}}\right]\rho_{n_{\alpha},n_{\beta}+1},
\label{eq:rho diag}
\end{eqnarray}
where
$K_{n_{\alpha},n_{\beta}} = M_{n_{\alpha},n_{\beta}} + ({\bar{\delta}A/2)^2 [1+
\bar{\delta}^{2}+(B/A)(n_{\alpha}+1/2)+(B/4A)^{2}}]^{-2}
M_{n_{\alpha},n_{\beta}}^{-1}$
and
$M_{n_{\alpha},n_{\beta}} = [A(n_{\alpha}+1/2)+B/4] [1 + \bar{\delta}^{2} + (B/A)(n_{\alpha}+1/2)+(B/4A)^{2}]^{-1}
+C_{1}(n_{\alpha}-1/2)+C_{2}(n_{\beta}-1/2)$.

Finally, the equations for the probability distribution of the number of photons in the modes $\alpha$ and $\beta$,
$p(n_{\alpha}) = \rho_{n_{\alpha}} = \sum_{n_{\beta}}\rho_{n_{\alpha},n_{\beta}}$ and
$p(n_{\beta}) = \rho_{n_{\beta}}
= \sum_{n_{\alpha}}\rho_{n_{\alpha},n_{\beta}}$, follow:
\begin{eqnarray}
&&\dot{p}(n_{\alpha})=\frac{An_{\alpha}}{1+\bar{\delta}^{2}+(B/A)n_{\alpha}}p(n_{\alpha}-1)-C_{1}n_{\alpha}p(n_{\alpha})-\frac{A(n_{\alpha}+1)}{1+\bar{\delta}^{2}+(B/A)(n_{\alpha}+1)}p(n_{\alpha})
\nonumber \\
&+&C_{1}(n_{\alpha}+1)p(n_{\alpha}+1)-2C_{3}^{2}n_{\alpha}\sum_{n_{\beta}}n_{\beta}K_{n_{\alpha},\, n_{\beta}}^{-1}\left[2p(n_{\alpha},n_{\beta})-p(n_{\alpha},n_{\beta}-1)-p(n_{\alpha}-1,n_{\beta})\right]
\nonumber \\
&+&2C_{3}^{2}(n_{\alpha}+1)\sum_{n_{\beta}}n_{\beta}K_{n_{\alpha}+1,\, n_{\beta}}^{-1}\left[2p(n_{\alpha}+1,n_{\beta})-p(n_{\alpha}+1,n_{\beta}-1)-p(n_{\alpha},n_{\beta})\right],
\label{eq:rho alpha}
\end{eqnarray}
\begin{eqnarray}
&&\dot{p}(n_{\beta})=-C_{2}n_{\beta}p(n_{\beta})+C_{2}(n_{\beta}+1)p(n_{\beta}+1)\nonumber \\
&-&2C_{3}^{2}n_{\beta}\sum_{n_{\alpha}}n_{\alpha}K_{n_{\alpha},\, n_{\beta}}^{-1}\left[2p(n_{\alpha},n_{\beta})-p(n_{\alpha}-1,n_{\beta})-p(n_{\alpha},n_{\beta}-1)\right]
\nonumber \\
&+&2C_{3}^{2}(n_{\beta}+1)\sum_{n_{\alpha}}n_{\alpha}K_{n_{\alpha},\, n_{\beta}+1}^{-1}\left[2p(n_{\alpha},n_{\beta}+1)-p(n_{\alpha}-1,n_{\beta}+1)-p(n_{\alpha},n_{\beta})\right].
\label{eq:rho beta}
\end{eqnarray}

In the steady-state regime, $\dot{p}(n_{\alpha}) = \dot{p}(n_{\beta}) = 0$ and Eqs. (\ref{eq:rho alpha}) and (\ref{eq:rho beta}) can be reduced to two-term recurrence relations by using the detailed balance condition and assuming that $\sum_{n_{j}}F(n_{i},n_{j})p(n_{i},n_{j}) \simeq F(n_{i},\bar{n}_{j})p(n_{i})$. Here $\bar n_{\alpha}$ and $\bar n_{\beta}$ denote the average photon numbers in the modes $\alpha$ and $\beta$, respectively. For $n_{\alpha}$, $n_{\beta} \geq 1$, the resulting equations are
\begin{eqnarray}
&&p(n_{\alpha}) \left\{
C_1 - 2 C_3^2 \left[ (\bar{n}_{\beta} + 1)
K_{n_{\alpha}, \bar{n}_{\beta}+1}^{-1} - 2 \bar{n}_{\beta}
K_{n_{\alpha}, \bar{n}_{\beta}}^{-1} \right] \right\}
\nonumber \\
&-& p(n_{\alpha}-1) \left(
\frac{A}{1 + \bar{\delta}^{2}+(B/A) n_{\alpha}}
+ 2 C_3^2 \bar{n}_{\beta}
K_{n_{\alpha}, \bar{n}_{\beta}}^{-1} \right) = 0,
\label{eq:mode alpha}
\\
&&p(n_{\beta}) \left\{
C_2 - 2 C_3^2 \left[ (\bar{n}_{\alpha} + 1)
K_{\bar{n}_{\alpha}+1, n_{\beta}}^{-1} - 2 \bar{n}_{\alpha}
K_{\bar{n}_{\alpha}, n_{\beta}}^{-1} \right] \right\}
\nonumber \\
&-& p(n_{\beta}-1)
\times 2 C_3^2 \bar{n}_{\alpha}
K_{\bar{n}_{\alpha}, n_{\beta}}^{-1} = 0.
\label{eq:mode beta}
\end{eqnarray}

The equations\ (\ref{eq:rho nondiag})--(\ref{eq:mode beta}) are the main result of this work. Supplemented by the normalization condition $\sum_{n_{\alpha}} p(n_{\alpha}) =\sum_{n_{\beta}} p(n_{\beta}) = 1$, they will allow us to analyze the photon statistics, the threshold, the photon number fluctuations and the linewidth of the open-cavity laser in the steady-state regime.

\subsection{Photon statistics}

Photon number distributions $p(n_{\alpha})$ and $p(n_{\beta})$ can be readily obtained by solving Eqs.\ (\ref{eq:mode alpha}) and (\ref{eq:mode beta}) numerically. But even without any numerical solution, it is easy to convince oneself that the only solution of Eq.\ (\ref{eq:mode beta}) is $p(n_{\beta}) = 0$ for $n_{\beta} \geq 1$ [and hence $p(n_{\beta} = 0) = 1$ by normalization]. Equation (\ref{eq:mode beta}) does not contain any gain, only damping terms are present. Besides, the terms proportional to $C_{3}$ cancel each other well above threshold, i.e. for $\bar{n}_{\alpha} \gg 1$, and therefore the steady-state solution vanishes, i.e. $\rho_{n,n}^{(\beta)}=0$, with the exception that $\rho_{0,0}^{(\beta)}=1$ involving the normalization condition.
In contrast, Eq.\ (\ref{eq:mode alpha}) does have a non-trivial solution $p(n_{\alpha}) > 0$ for $n_{\alpha} \geq 1$ and this solution depends on the pump rate $r$. We therefore expect that if the laser effect occurs in our system, we should look for it signatures in the behavior of the composite mode $\alpha$.

In the limit of weak pump $r \rightarrow 0$, we may consider the linear approximation for the laser equations, i.e. $B = 0$, and the photon number distribution resulting from Eq.\ (\ref{eq:mode alpha}) approaches the thermal distribution as we see in Fig. 1. The analytical solution of  Eq.\ (\ref{eq:mode alpha}) (with $\bar{n}_{\beta} \simeq 0$) can be approximated by
\begin{eqnarray}
p(n_{\alpha}) \simeq \left(1-\frac{A}{\widetilde{C}_{1}}\right) \left(\frac{A}{\widetilde{C}_{1}} \right)^{n_{\alpha}},
\label{weakpump}
\end{eqnarray}
with $\widetilde{C}_{1}=C_{1}(1 + \bar{\delta}^{2})$, which is similar to the standard result for the single-mode laser \cite{Scully}.

\begin{figure}[t]
\includegraphics[width=0.9\textwidth]{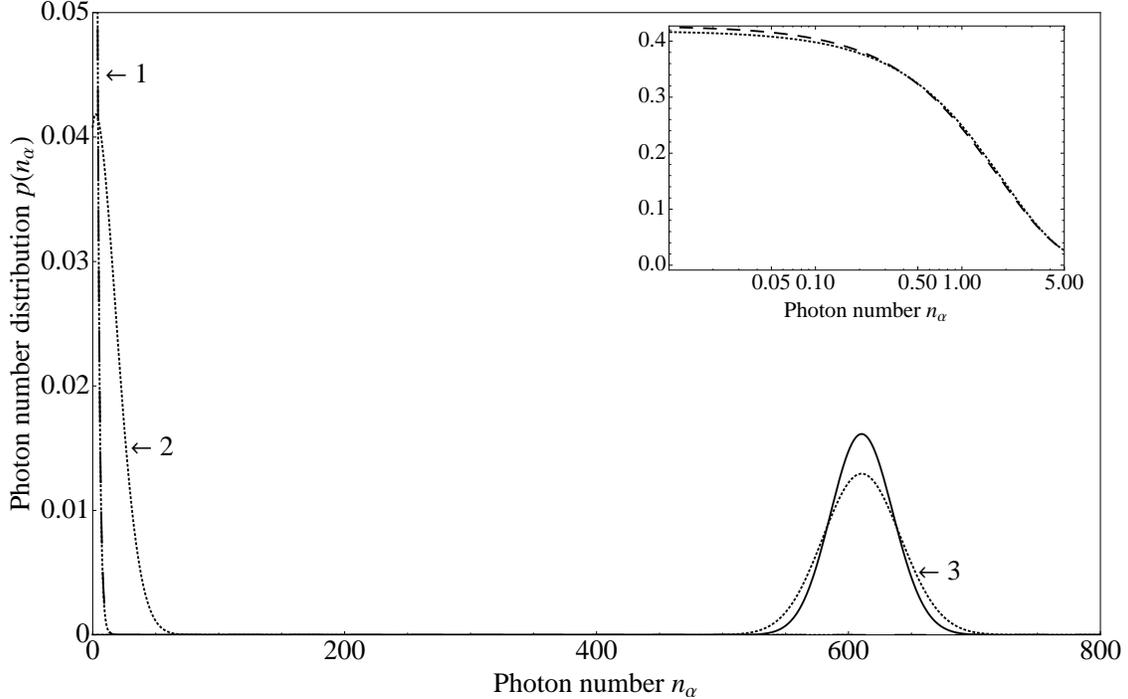}
\caption{Steady-state photon statistics for the composite mode $\alpha$ (dotted curve) below (curve 1), at (curve 2), and above (curve 3) threshold. The dashed and solid lines show the thermal and the Poisson distributions corresponding to the same average photon numbers $\bar{n}_{\alpha}$ as curves 1 and 3, respectively. For this figure, we fixed
$g_{1}/\Gamma=0.05$, $g_{2}/\Gamma=0.07$, $\delta/\Gamma=3$, $\gamma_{11}/\Gamma=6$,
$\gamma_{22}/\Gamma=5$, and $\gamma_{12}/\Gamma=5.5$.
\label{photonstatistics}}
\end{figure}

In the limit of strong pump, $r \rightarrow \infty$, saturation effects become important and $B\bar{n}_{\alpha}/A \gg 1+\bar {\delta}^{2}$. The analytical solution of Eq.\ (\ref{eq:mode alpha}) tends to
\begin{eqnarray}
p(n_{\alpha}) \simeq p(0)\frac{(\widetilde{A}/B)!(A^{2}/BC_{1})^{n_{\alpha}}}{(n_{\alpha}+\widetilde{A}/B)!},
\label{strongpump}
\end{eqnarray}
where $\widetilde{A}=A(1 + \bar{\delta}^{2})$, and $p(0)$ can be determined from the normalization of $p(n_{\alpha})$. We thus observe that the distribution of $n_{\alpha}$ changes qualitatively when the pump is increased and that its limiting forms (\ref{weakpump}) and (\ref{strongpump}) coincide with those for the single-mode laser \cite{Scully}. This suggests that the laser transition occurs for the composite mode $\alpha$ in our two-mode model. This is illustrated by the distributions of the photon number $n_{\alpha}$ found by solving Eqs.\ (\ref{eq:mode alpha}) and (\ref{eq:mode beta}) numerically that we show in Fig.\ \ref{photonstatistics}.
We see that at low pump (below threshold), the distribution is close to the thermal one: $p(n_{\alpha})=\left(1-\exp[-\hbar \bar{\omega} /k_{B}T] \right) \exp[-n_{\alpha} \hbar \bar{\omega }/k_{B}T]$, where the effective temperature $T$ is determined by $\exp[-\hbar \bar{\omega} /k_{B}T]=A/\widetilde{C}_{1}$. At strong pump (above threshold), $p(n_{\alpha})$ approaches the Poisson distribution: $p(n_{\alpha})=\exp[-\bar{n}_{\alpha}] \bar{n}_{\alpha}^{n_{\alpha}}/(n_{\alpha}!)$, where $\bar{n}_{\alpha}$ is given by Eq.\ (\ref{eq:phot no alpha}) below.

\subsection{Average photon number}

The average photon number $\bar{n}_{\alpha}$ can be found from Eq.\ (\ref{eq:mode alpha}) as
$\bar{n}_{\alpha} = \sum_{n_{\alpha}} n_{\alpha} p(n_{\alpha})$. Far above threshold, the distribution of $n_{\alpha}$ is strongly peaked around $\bar{n}_{\alpha}$ and $p(\bar{n}_{\alpha} + 1) \simeq p(\bar{n}_{\alpha})$, as well as  $K_{\bar{n}_{\alpha}+1,\,\bar{n}_{\beta}} \simeq K_{\bar{n}_{\alpha},\,\bar{n}_{\beta}+1} \simeq K_{\bar{n}_{\alpha},\,\bar{n}_{\beta}}$. Together with $\bar{n}_{\beta} = 0$ this yields
\begin{eqnarray}
\bar{n}_{\alpha} &\simeq& \frac{\widetilde{A}}{B}\left(\frac{A}{\widetilde{C}_{1}}-1\right).
\label{eq:phot no alpha}
\end{eqnarray}
Hence, the threshold for the composite mode $\alpha$ is given by the condition $A/C_{1} = 1 + \bar{\delta}^{2}$.

\begin{figure}[t]
\includegraphics[width=0.9\textwidth]{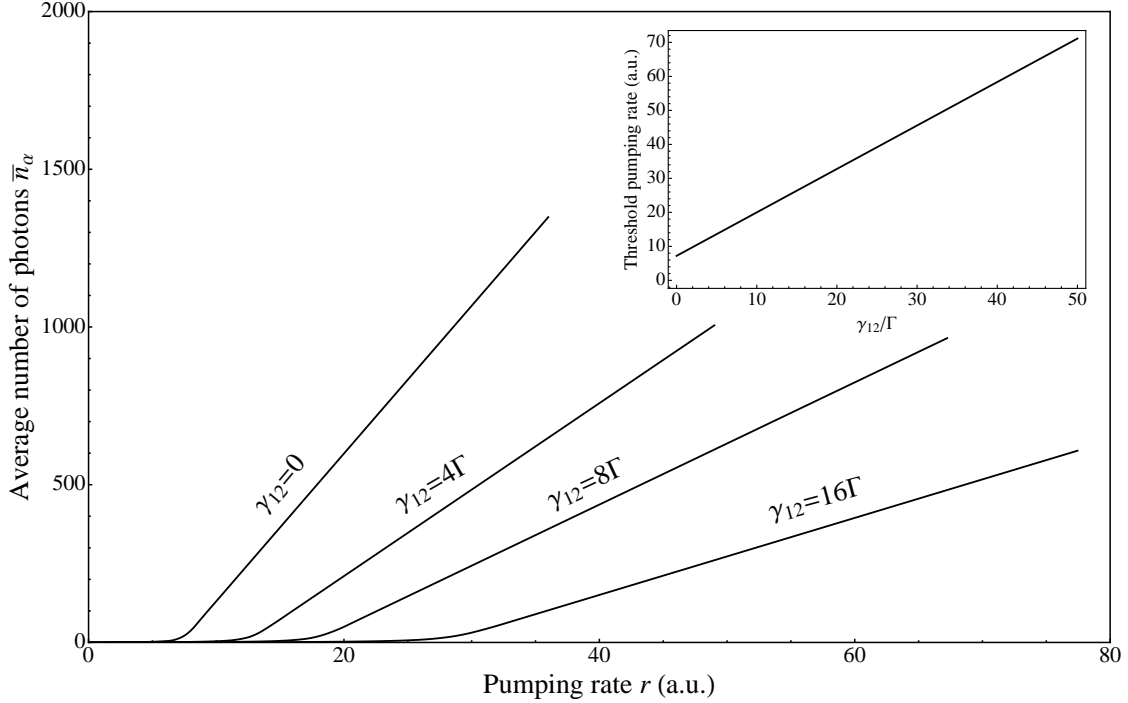}
\caption{
The average number of photons in the composite mode $\alpha$ as a function of the pumping rate $r$. The off-diagonal elements of the symmetric matrix $\gamma$ for the four curves are $\gamma_{12} = 0$, $4\Gamma$, $8\Gamma$ and $16\Gamma$, respectively. Other parameters are as in Fig.\ \ref{photonstatistics}. The inset shows the dependence of the threshold on $\gamma_{12}/\Gamma$.
\label{averagen}}
\end{figure}

The dependence of the threshold pumping rate $r$ on the off-diagonal element $\gamma_{12}$ of the damping matrix $\gamma$ is shown in the inset of Fig.\ \ref{averagen}. In contrast, the mode $\beta$ does not have a threshold and the number of photons in it is always equal to zero. The full dependence of $\bar{n}_{\alpha}$ on the parameters of the problem can be obtained by solving Eq.\ (\ref{eq:mode alpha}) numerically. In Fig.\ \ref{averagen} we show the dependence of $\bar{n}_{\alpha}$ on the pumping rate $r$ for different values of the off-diagonal element $\gamma_{12}$ of the damping matrix $\gamma$. Figures \ref{photonstatistics} and \ref{averagen} show that the composite mode $\alpha$ behaves as a lasing mode with a well-defined threshold. This is also highlighted by the formal equivalence of Eq.\ (\ref{eq:phot no alpha}) and the standard expression for the average photon number in a single-mode laser \cite{Scully,orszag08}.

\subsection{Photon number fluctuations}

A common way to characterize fluctuations of the photon number in a mode of the electromagnetic field is to compute the so-called Mandel parameter,
\begin{equation}
Q = \frac{\overline{n^{2}} - \bar{n}^{2}}{\bar{n}} - 1.
\label{eq:Mandel par}
\end{equation}
\begin{figure}[t]
\includegraphics[width=0.9\textwidth]{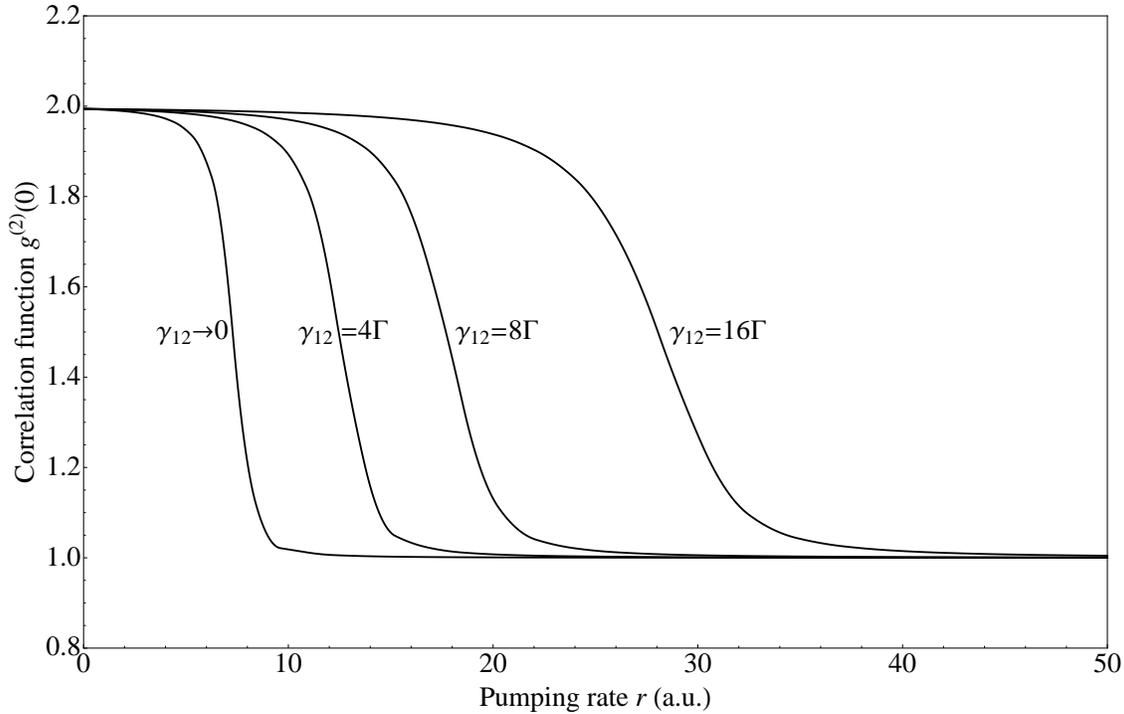}
\caption{The second-order correlation function $g^{(2)}(0)$ as a function of the pumping rate, for the same parameters as in Fig.\ \ref{averagen}.
\label{g2}}
\end{figure}
Equations (\ref{weakpump}) and (\ref{strongpump}) readily allow us to compute this quantity for the composite mode $\alpha$ analytically well below and  far above threshold, respectively, as
\begin{eqnarray}
Q_{\alpha} &\simeq&\frac {A}{\widetilde{C}_{1}-A}, \mbox{\;\;below threshold},
\label{qbelow}
\\
Q_{\alpha} &\simeq& \frac {\widetilde{C}_{1}}{A-\widetilde{C}_{1}}, \mbox{\;\;above threshold}.
\label{qabove}
\end{eqnarray}

Another quantity which is often used to characterize fluctuations of the photon number in experiments
(see, e.g., Ref.\ \cite{Wiersig}) is the second-order correlation function
$g^{(2)}(0) = \langle \hat{\alpha}^{\dagger} \hat{\alpha}^{\dagger}
\hat{\alpha} \hat{\alpha} \rangle/
\langle \hat{\alpha}^{\dagger} \hat{\alpha} \rangle^2 = Q_{\alpha}/\bar{n} + 1$. Its value ranges from $g^{(2)}(0) = 2$ for thermal light to $g^{(2)}(0) = 1$ for coherent laser light. Hence, the dependence of $g^{(2)}(0)$ on the pumping rate shown in Fig.\ \ref{g2} allows one to identify the laser transition quite clearly.

\subsection{Laser frequency and linewidth}

Information about the frequency and the linewidth of light emitted by the two-mode open-cavity laser can be extracted from the off-diagonal elements of the density matrix, Eq.\ (\ref{eq:rho nondiag}) \cite{Scully}.  Defining
$\rho_{n_{\alpha},n_{\beta}}(k_{1},k_{2}) = \rho_{n_{\alpha},n_{\beta};\, n_{\alpha}+k_{1},n_{\beta}+k_{2}}$, we follow Ref.\ \cite{Swain} and use an ansatz $\dot{\rho}_{n_{\alpha},n_{\beta}}(k_{1},k_{2})=
-\mu(k_{1},k_{2})\rho_{n_{\alpha},n_{\beta}}(k_{1},k_{2})$. When we insert this into Eq.\ (\ref{eq:rho nondiag}), it follows that, up to the lowest non-vanishing order in $k_1 k_2$ (valid for  $k \ll n$),
\begin{align}
\mu(k_{1},k_{2}) & \simeq\frac{k_{1}^{2}}{8}\left(\frac{A/(\bar{n}_{\alpha}+1)+2B}{1+\bar{\delta}^{2}+(B/A)(\bar{n}_{\alpha}+1+k_{1}/2)+(B/4A)^{2}k_{1}^{2}}+\frac{C_{1}}{\bar{n}_{\alpha}}\right)+\frac{k_{2}^{2}C_{2}}{8\bar{n}_{\beta}}\nonumber \\
& -\frac{iA\bar{\delta}k_{1}/2}{1+\bar{\delta}^{2}+(B/A)(\bar{n}_{\alpha}+1+k_{1}/2)+(B/4A)^{2}k_{1}^{2}}.\label{eq:miu k1k2}
\end{align}

\begin{figure}[t]
\includegraphics[width=0.9\textwidth]{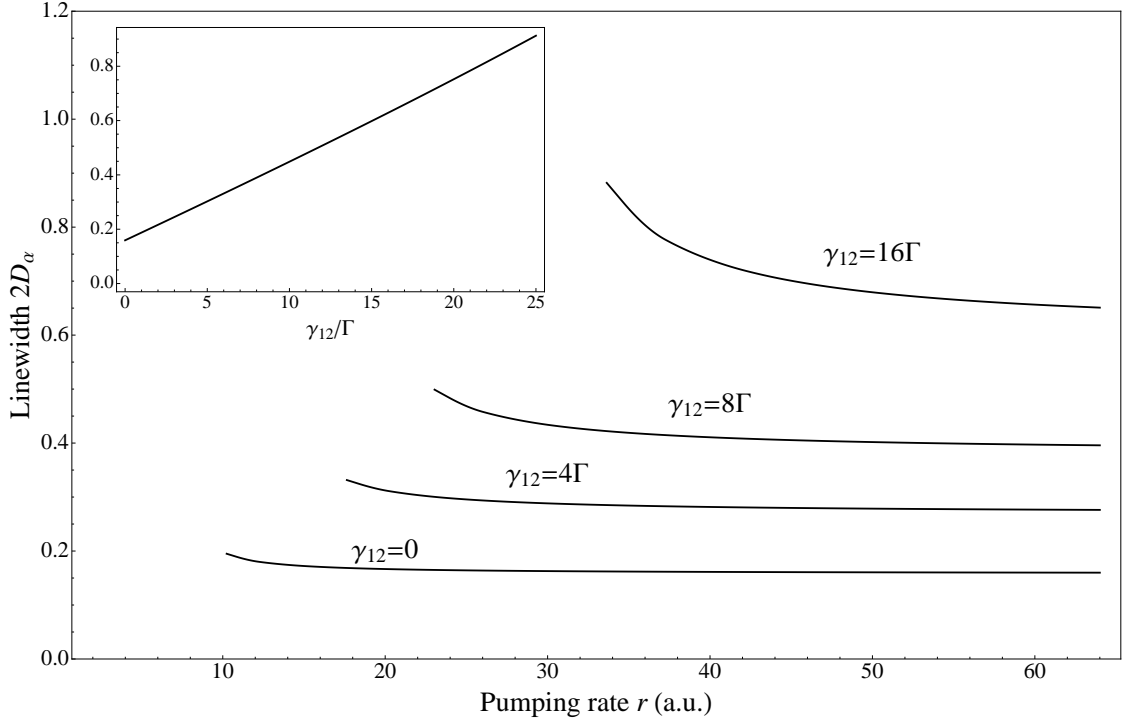}
\caption{Linewidth of the composite mode $\alpha$ as a function of the pumping rate $r$ for the same parameters as in Fig.\ \ref{averagen}. The inset shows the dependence of the linewidth on $\gamma_{12}/\Gamma$ for $r = 100$.
\label{linewidth}}
\end{figure}

The linewidth of the laser emission corresponding to the composite mode $\alpha$ is given by the real part of $\mu(1,0)$,
\begin{equation}
2D_{\alpha} =
\frac{1}{4}\left[\frac{A/(\bar{n}_{\alpha}+1)+
2B}{1+\bar{\delta}^{2}+(B/A)(\bar{n}_{\alpha}+3/2)+
(B/4A)^{2}}+\frac{C_{1}}{\bar{n}_{\alpha}}\right].
\label{eq:linewidth D}
\end{equation}
For $B/A \ll 1$ and $\bar{\delta} = 0$ the linewidth reduces to  $2D_{\alpha} = (A+C_{1})/4\bar{n}_{\alpha}$. This formally coincides with the result of the standard laser theory \cite{Scully}, except for the definition of $C_1$, which includes an additional term, $\propto \gamma_{12}$, in our case of the open-cavity laser.
In Fig.\ \ref{linewidth}, we plot the dependence of the linewidth of mode $\alpha$ on the pumping rate $r$ for $r$ at least 25\% above threshold. The dependence of the linewidth on the off-diagonal element of the damping matrix $\gamma$ is shown in the inset for $r = 100$, which is far above threshold for the range of $\gamma_{12}$ shown in the figure.

\begin{figure}[t]
\includegraphics[width=0.9\textwidth]{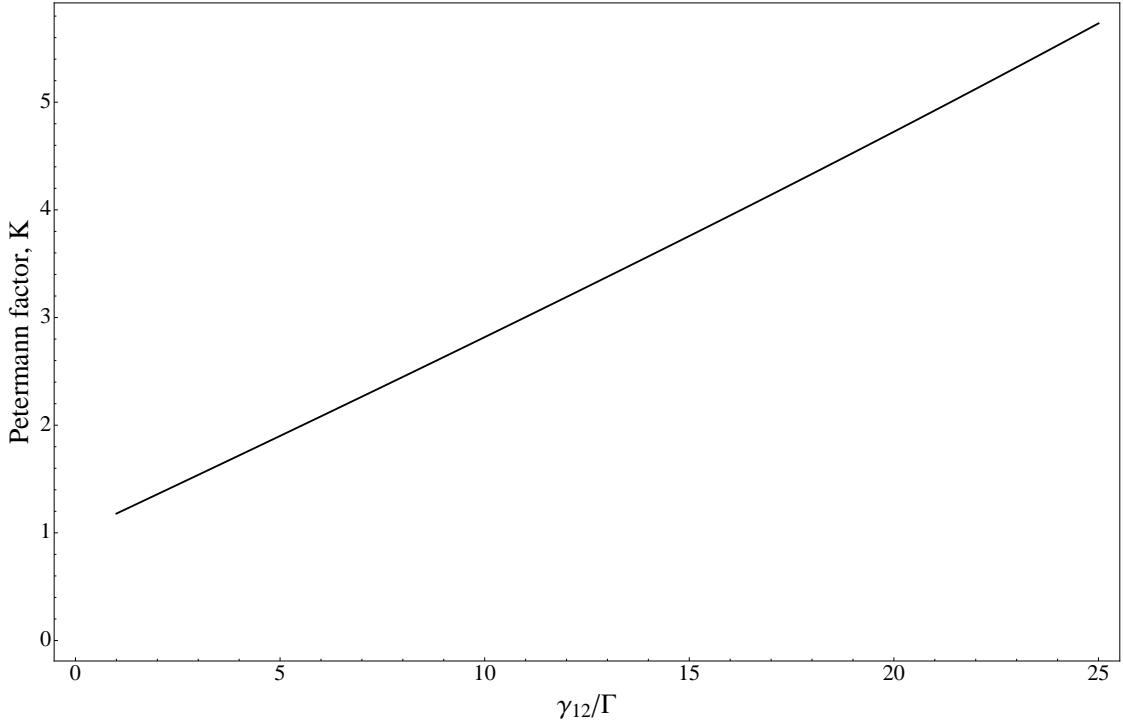}
\caption{The Petermann factor $K$ as a function of $\gamma_{12}/ \Gamma$ for the same values of parameters as in Fig.\ \ref{averagen} and large pumping rate $r = 100$.
\label{Kfactor}}
\end{figure}

As in the usual single-mode laser with a cavity of high
quality factor \cite{siegman86,Scully}, the finite linewidth of our two-mode open-cavity laser is due to the spontaneous emission of the active atoms. Previous studies have shown that in an open cavity, the emission line is further broadened by a factor $K$, called the Petermann factor \cite{Petermann}, due to the nonorthogonality of the cavity modes \cite{patra00}. However, our analysis here is based on the quantization procedure of Refs.\ \cite{hackenbroich02} that relies on the expansion of the electromagnetic field inside and outside of the cavity in terms of orthogonal modes. It is therefore interesting to check if our model reproduces the large Petermann factor expected from previous studies \cite{patra00}. We define
\begin{equation}
K=\frac{D_{\alpha}(\gamma_{12})}{D_{\alpha}(0)},
\label{eq:Kfactor}
\end{equation}
where the linewidth $D_{\alpha}$ is considered as a function of $\gamma_{12}$ that quantifies the openness of the system in our model. Already from Fig.\ \ref{linewidth} we see that $D_{\alpha}$ increases with $\gamma_{12}$. This is confirmed by Fig.\ \ref{Kfactor} where $K$ is shown as a function of $\gamma_{12}$ far above threshold.
A simple analytical expression for $K$ can be obtained at
$B/A\ll1$ and for $\delta=0$:
\begin{equation}
K=\frac{C_1(\gamma_{12})\left[A+C_1(\gamma_{12})\right] \left[A-C_1(0) \right]}{C_1(0) \left[A-C_1(\gamma_{12}) \right] \left[A+C_1(0) \right]}.
\label{eq:Kasymp}
\end{equation}
Because $C_1(\gamma_{12}) \propto \gamma_{12}$, we find that the growth of $K$ with $\gamma_{12}$ is roughly linear, which is in agreement with Fig.\ \ref{Kfactor}. Thus, our model reproduces the increased Petermann factor in open laser systems, despite the orthogonality of the basis in which we quantized the electromagnetic field.

Finally, the imaginary part of $\mu(1, 0)$ yields the shift of the laser frequency with respect to $\bar{\omega}$:
\begin{equation}
\Delta_{\alpha} = -\frac{A\bar{\delta}/2}{1+\bar{\delta}^{2}+
(B/A)(\bar{n}_{\alpha}+3/2)+(B/4A)^{2}}.
\label{eq:lasing freq}
\end{equation}
This equation shows that the frequency shift depends on the average photon number $\bar{n}_\alpha$ which, in turn, is a function of the off-diagonal element of the damping matrix $\gamma$.

\section{Conclusion}

We developed the quantum theory of a laser with an open cavity. The openness of the cavity is mathematically described by a nondiagonal damping matrix $\gamma$. Assuming that only two modes of the ``cold'' cavity are allowed to participate in the competition for gain, we have shown that the modes strongly interact with each other and that a composite mode (denoted by $\alpha$ here) is built up. This composite mode shows all the properties of a typical laser mode: threshold behavior, photon statistics evolving from thermal to Poissonian as the pumping rate increases, augmented photon number fluctuations in the vicinity of the threshold, and linewidth narrowing. At the same time, the precise behavior of the composite mode $\alpha$ at given values of parameters depends explicitly on the off-diagonal element $\gamma_{12}$ of the damping matrix $\gamma$. More precisely, an increase of $\gamma_{12}$ rises the lasing threshold and broadens the laser emission line.

An important result that cannot be obtained from a semi-classical theory and needs the quantum theory developed in this paper to be understood is the broadening of the emission linewidth due to the coupling of the cavity modes through a common bath (Fig.\ \ref{Kfactor}). This broadening should be accessible experimentally: indeed, measurements of the laser linewidth in a laser with two coupled modes were already performed in, e.g., Ref.\ \cite{Steiner}. It is interesting to note that in that work the mechanism of coupling was different and led to narrowing of the emission linewidth due to the so-called correlated spontaneous emission in a system of two excited levels that provide inversion for lasing on transitions sharing a common lower level (see Ref.\ \cite{Bergou} for the theoretical model). In contrast to this, the two-mode open-cavity laser considered in this paper exhibits widening of the emission linewidth. This would certainly limit applications of such a laser for high-precision measurements but might be beneficial for other applications where low-coherence light is required (like, e.g., optical coherence tomography \cite{fercher03}).

One of the possible applications of our theory may lie in the field of random lasers. In this case, the damping matrix $\gamma$ should be treated as a random matrix and our results should be averaged over the statistical distribution of its elements. However, to obtain results that can be directly applied to random laser systems, one has to generalize our analysis to the multi-mode case because the number of active modes in random lasers is expected to be large \cite{hackenbroich05}.

\acknowledgements

This work was partially supported by Chilean projects Mecesup (no. SSM0605) and Fondecyt (no. 1100039), as well as by the French ANR (project no. 06-BLAN-0096 CAROL).

\end{document}